\documentclass{pasj00}

\begin{document}
\SetRunningHead{S.\ Yamauchi et al.}{X-ray Filament with a Strong 6.7 keV Line in the Galactic Center Region}

\title{X-ray Filament with a Strong 6.7 keV Line in the Galactic Center Region}

\author{
Shigeo \textsc{Yamauchi}\altaffilmark{1},
Miku \textsc{Shimizu}\altaffilmark{1},
Shinya \textsc{Nakashima}\altaffilmark{2},
Masayoshi \textsc{Nobukawa}\altaffilmark{3, 4},
Takeshi Go \textsc{Tsuru}\altaffilmark{3},
and 
Katsuji \textsc{Koyama}\altaffilmark{3,5}}
 \altaffiltext{1}{Department of Physics, Nara Women's University, Kitauoyanishimachi, Nara 630-8506}
\email{yamauchi@cc.nara-wu.ac.jp}
\altaffiltext{2}{Institute of Space and Astronautical Science/JAXA, 
  3-1-1 Yoshinodai, Chuo-ku, Sagamihara, Kanagawa 252-5210}
\altaffiltext{3}{Department of Physics, Graduate School of Science, Kyoto University, \\
Kitashirakawa-oiwake-cho, Sakyo-ku, Kyoto 606-8502}
\altaffiltext{4}{The Hakubi Center for Advanced Research, Kyoto University, Yoshida-Ushinomiya-cho, Kyoto 606-8302, Japan}
\altaffiltext{5}{Department of Earth and Space Science, Graduate School of Science, Osaka University, \\
1-1 Machikaneyama-cho, Toyonaka, Osaka 560-0043}


\maketitle

\begin{abstract}

An elongated  X-ray source with a strong K-shell line from He-like iron (Fe XXVI) is found at (RA, Dec)$_{\rm J2000.0}$=(\timeform{17h44m00s.0},
 \timeform{-29D13'40''.9}) in the Galactic center region.  
The position coincides with the X-ray thread, G359.55$+$0.16, which is aligned with the radio non-thermal filament.
The X-ray spectrum is well fitted with an absorbed thin thermal plasma ({\tt apec}) model. 
The best-fit temperature, metal abundance, and column density are 4.1$^{+2.7}_{-1.8}$ keV, 0.58$^{+0.41}_{-0.32}$ solar, 
and 6.1$^{+2.5}_{-1.3}$$\times$10$^{22}$ cm$^{-2}$, respectively.
These values are similar to those of the largely extended Galactic center X-ray emission.

\end{abstract}

\section{Introduction}

The central 300 pc of the Milky Way Galaxy (hereafter, the Galactic center: GC) is a unique region that contains many objects associated 
with high-energy phenomena.
Due to its proximity, the GC would be one of the best region to investigate 
complex activity possibly originated from the central supermassive black hole.

Sagittarius (Sgr) A$^{\ast}$, a bright compact radio source located at the dynamical center of the Galaxy, 
is a supermassive black hole of 4$\times$10$^{6}$ $M_{\odot}$
(e.g., \cite{Genzel2000,Ghez2000,Schoudel2002,Eckart2002}).
Other than Sgr A$^{\ast}$, the most striking structure is  the "radio arc", consisting of 
a straight filament (called the spur) perpendicular to the Galactic plane
and arched filament (called the bridge) along the Galactic plane
(e.g., \cite{Genzel1987,Genzel1994}, and references therein).
Previous radio observations revealed that the bridge is thermal while the spur is non-thermal origins.
In addition, a number of non-thermal radio filaments (NTFs) that run roughly aligned perpendicular to the Galactic plane have been found 
(e.g., \cite{Yusef1984}). 
The NTFs are due to synchrotron radiation from relativistic particles on local magnetic field.

 Sgr A$^{\ast}$ is not  X-ray bright; its luminosity is only $\sim$10$^{33}$ erg s$^{-1}$ (e.g., \cite{Baganoff2003}) 
 with sporadic small flares up to $\sim$10$^{35}$ erg s$^{-1}$ (e.g., \cite{Baganoff2001}).  
 However the discovery of time variable  6.4 keV  (K-shell emission from Fe I) clouds indicates that Sgr A$^{\ast}$ had been more active of 
 about  $\sim$10$^{39-40}$ erg s$^{-1}$, till about 50 years ago (e.g., \cite{Koyama1996,Ponti2010,Ryu2013}).

\begin{figure*}
  \begin{center}
        \includegraphics[width=16cm]{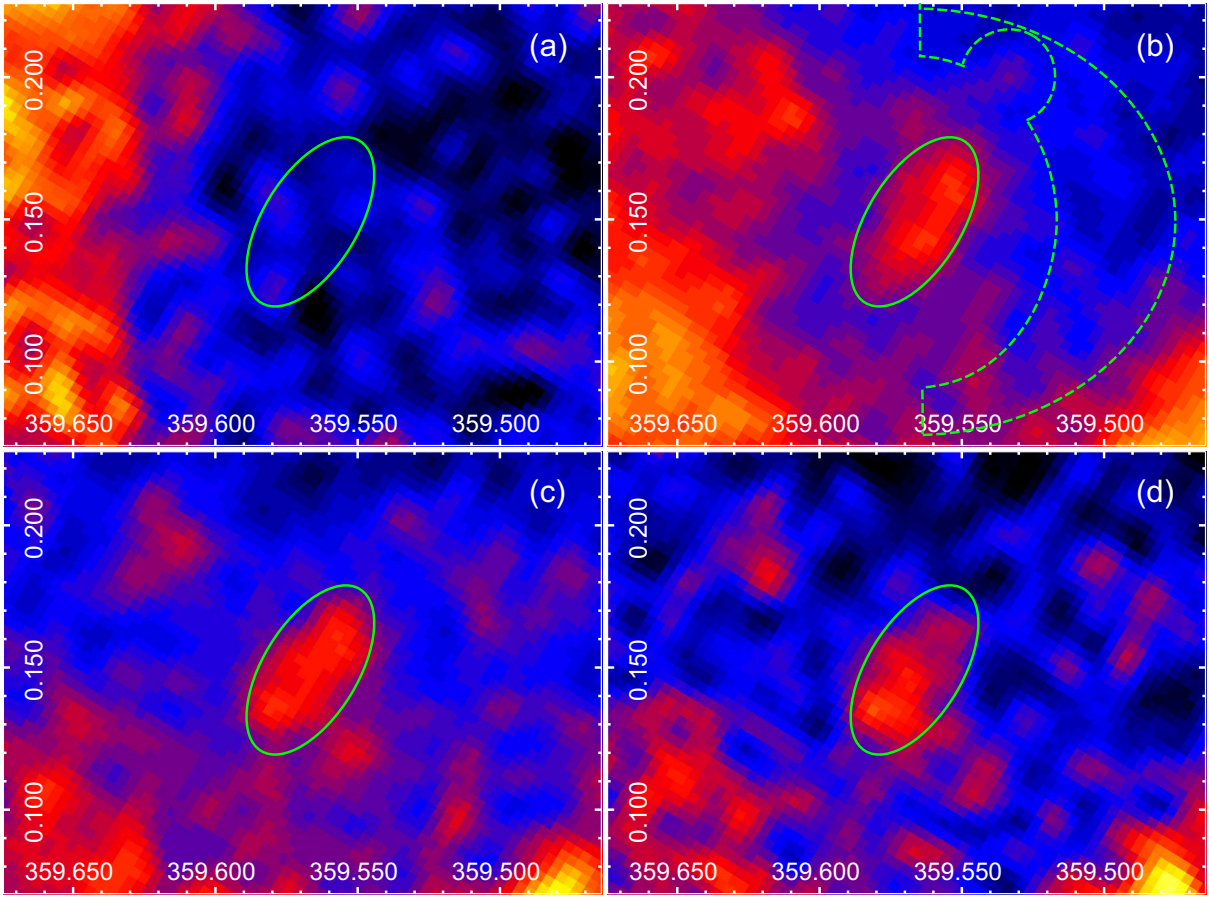}
  \end{center}
\caption{XIS images in the (a) 1--2, (b) 2--5, and  (c) 5--8, (d) 6.55--6.80 keV (Fe XXV-K$\alpha$).
The coordinates are Galactic.  
All the available data acquired in the Galactic center region are utilized.
Subtraction of the non-X-ray background and vignetting correction are performed.
The green ellipses in all panels show a position of Suzaku J174400$-$2913 (a source region, see text),
while the dashed line in (b) shows a background region. 
}\label{fig:sample}
\end{figure*}

At present, the most  prominent X-ray object in the GC region is 
a thin hot plasma with a temperature of $\sim$7 keV, 
the so-called Galactic Center X-ray Emission (GCXE) (Koyama et al. 1989, 1996, 2007c, Yamauchi et al. 1990). 
The total thermal energy of this plasma is huge, ($\sim$10$^{53-54}$ erg, \cite{Yamauchi1990,Koyama1996}), 
and the plasma temperature is too high to confine in the GC against the Galactic gravity. Thus, its origin is still puzzling.
In addition to the GCXE, 
some filamentary X-ray structures were discovered with Chandra and XMM-Newton in the 
GC region (Wang et al. 2002; Sakano et al. 2003; Lu et al. 2003, 2008; Johnson et al. 2009).
Most of them exhibit featureless spectra with no sign of emission lines from highly ionized atoms, and hence they are likely to be in non-thermal nature.

G359.55$+$0.16 is one of the X-ray filament discovered with Chandra, 
and is named  ''X-ray thread'' \citep{Wang2002,Lu2003,Johnson2009}.
This source would have a cylindrical shape of 10$''$ diameter and 1--2$'$ length, aligned with the NTF G359.54$+$0.18 \citep{Lu2003,Yusef1997}.
\citet{Lu2003} concluded that G359.55$+$0.16 is likely to be a non-thermal emission.
In fact, \citet{Johnson2009} reported 
that the X-ray spectrum is well represented by a power-law model with a photon index of 
1.2$^{+1.3}_{-1.1}$ and absorption with $N_{\rm H}$=5.5$^{+3.9}_{-3.6}\times$10$^{22}$ cm$^{-2}$. 
The energy flux in the 2--10 keV band was estimated to be 1.8$\times$10$^{-13}$ erg s$^{-1}$ cm$^{-2}$.

The Suzaku XIS has a better spectral resolution and lower/more stable intrinsic background
than those of the previous X-ray satellites \citep{Koyama2007a,Mitsuda2007}, and hence 
it has the best sensitivity especially in the Fe K-shell line band.
We carried out survey observations in the GC region with Suzaku and
found an elongated  emission with a strong Fe-K emission line (6.7 keV)
at the position of G359.55$+$0.16. 
This paper reports the results of the spectral analysis of this elongated emission.
Throughout this paper, the quoted errors are at the 90\% confidence level.

\section{Observation and Data Reduction}

Survey observations in the GC region were carried out with the XIS \citep{Koyama2007a} 
onboard Suzaku \citep{Mitsuda2007}.
The XIS is composed of 4 CCD camera sensors placed at the focal planes of the thin foil X-ray Telescopes (XRT, \cite{Serlemitsos2007}). 
XIS\,1 is a back-side illuminated (BI) CCD, while XIS\,0, 2, and 3 are front-side illuminated (FI) CCDs. 
The Field of view (FOV) of the XIS is \timeform{17'.8}$\times$\timeform{17'.8}.
One of the FI sensors (XIS\,2) became non-functional since 2006. 
Therefore, we used the data obtained with the other three sensors (XIS\,0, 1, and 3). 
The XIS was operated in the normal clocking mode with the time resolution of 8 s.
The spectral resolution of the XIS was degraded due to the radiation of cosmic particles.
In order to restore the XIS performance, the spaced-row charge injection (SCI) technique was applied.
Details of the SCI technique are given in \citet{Nakajima2008} and \citet{Uchiyama2009}.

The northwest of the GC field was observed on 2009 March 6 2:39:12 -- 
March 9 2:55:25 (Obs. ID 503072010).
The center position was 
($\alpha$, $\delta$)$_{\rm J2000.0}$=(\timeform{265D.9883}, \timeform{-29D.2111}) 
[($l$, $b$)=(\timeform{359D.5753}, \timeform{+0D.1669})].

Data reduction and analysis were made with the HEAsoft version 6.13. 
The XIS pulse-height data for each X-ray event were converted to 
Pulse Invariant (PI) channels using the {\tt xispi} software 
and the calibration database version 2013-03-05.
We rejected the data taken at the South Atlantic Anomaly, 
during the earth occultation, and at the low elevation angle 
from the earth rim of $<5^{\circ}$ (night earth) and $<20^{\circ}$ (day earth).  
After these screening, the exposure time was 140.6 ks. 

\section{Analysis and Results}

\subsection{Image}
Figures 1a, 1b, 1c and 1d show the X-ray images in the 1--2, 2--5, 5--8, and 6.55--6.80 keV (Fe XXV-K$\alpha$) energy bands, respectively.
The data of XIS 0, 1, and 3 were added to increase photon statistics.
An elongated emission is found above 2 keV (figures 1b, 1c, and 1d), but no emission is seen below 2 keV (figure 1a).  
The center position of the elongated emission was determined to be 
(RA, Dec)$_{\rm J2000.0}$=(\timeform{17h44m00s.0}, \timeform{-29D13'40''.9})
[  ($l$, $b$)=(\timeform{359D.566}, \timeform{+0D.149}) ],
and hence we named this source Suzaku J174400$-$2913.
Comparing with the Chandra image, we found that Suzaku J174400$-$2913 
coincides with the "X-ray thread" G359.55$+$0.16 \citep{Wang2002,Lu2003,Johnson2009}.

\subsection{Spectrum}
X-ray spectra of Suzaku J174400$-$2913
were extracted from an ellipse with the major and  minor radii of 2$'$ and 1$'$, respectively.
Taking account of the large $b$-dependence of the GCXE flux (e.g., \cite{Yamauchi1990,Uchiyama2013}),  
we extracted the background spectra from a nearby source-free region with the same $b$ as the source.
The source and the background regions are shown by the solid and dotted lines,
respectively in figure 1.

We constructed the non-X-ray background (NXB) for the source and the background spectra
from the night-earth data (version 2013-06-01) using {\tt xisnxbgen} \citep{Tawa2008}.
For both the NXB-subtracted source and background spectra, 
we made vignetting correction according to the method shown by \citet{Hyodo2008},
and subtracted the background spectra from the source spectra. 
The source count rates after the background subtraction were (4.5$\pm$0.7)$\times$10$^{-3}$,  (4.5$\pm$0.7)$\times$10$^{-3}$,  
and (5.2$\pm$0.7)$\times$10$^{-3}$ counts s$^{-1}$ in the 2--8 keV band for XIS 0, XIS 1, and XIS 3, respectively. 
Then we merged the XIS 0 and XIS 3 spectra, because the response functions of the FIs are essentially the same with each other. 
The background-subtracted source spectra are shown in figure 2. 
In the spectra, we see strong emission lines at $\sim6.7$ keV. 

\begin{figure}[t]
  \begin{center}
       \includegraphics[width=8cm]{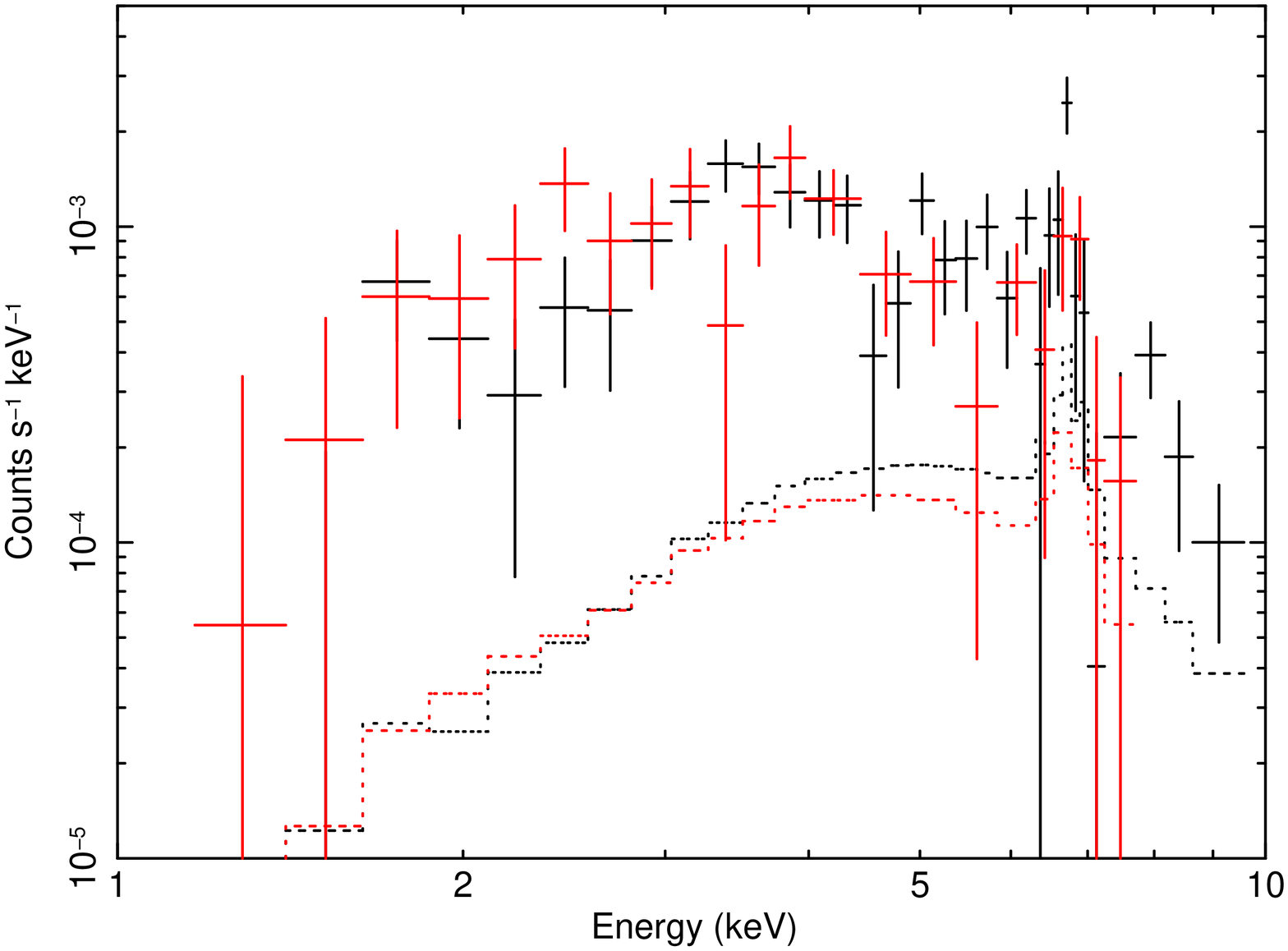}
   \end{center}
  \caption{
X-ray spectra of Suzaku J174400$-$2913 (black: XIS 0$+$3 and red: XIS 1).
The dotted lines show the contribution of the residual Chandra point sources.
}\label{fig:sample}
\end{figure}

Since Chandra resolved many point sources in the background and source regions (e.g., \cite{Muno2004, Muno2009}),  
we examined the contribution of the point sources. 
The source region of Suzaku J174400$-$2913 contains 11 Chandra sources. 
The integrated photon flux of the point sources is (1.41$\pm$0.21)$\times$10$^{-5}$ 
photons s$^{-1}$ cm$^{-2}$ in the 2--8 keV energy band (the Chandra GC source catalog, \cite{Muno2009}).
The background region contains 21 sources, then the integrated photon flux is (2.25$\pm$0.22)$\times$10$^{-5}$ photons s$^{-1}$ cm$^{-2}$ 
in the 2--8 keV band. 
Since the area of the background region is about 3 times larger than the source region, 
7.0$\times$10$^{-6}$  photons s$^{-1}$ cm$^{-2}$ (2--8 keV) should be remained in the source spectra, 
if a simple background subtraction is made (only difference of area is corrected).  
Thus we must estimate the remaining point source spectra for the spectral fitting by making a model spectrum of the Chandra point sources.

\begin{figure*}[t]
\begin{center}
\includegraphics[width=8cm]{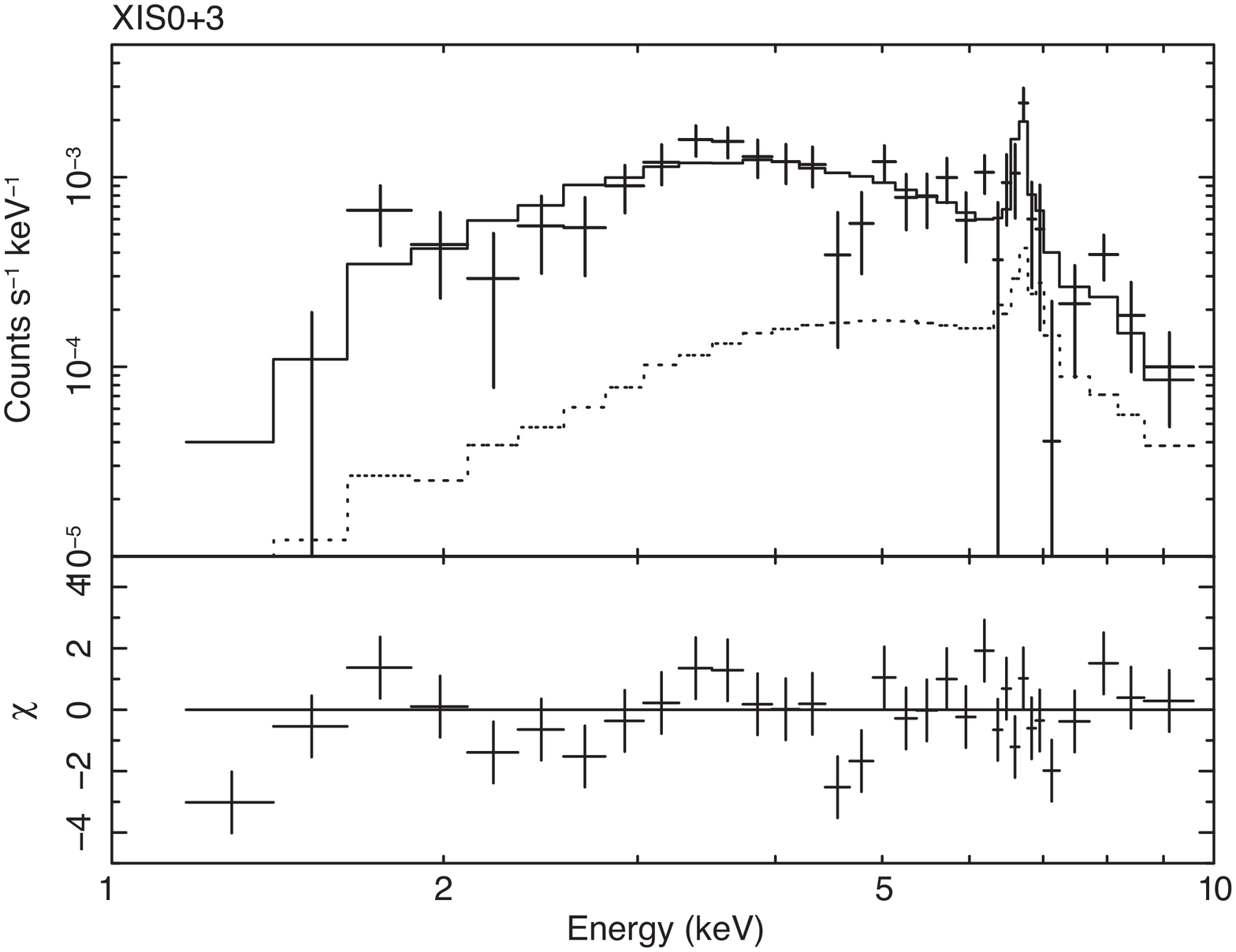}
\includegraphics[width=8cm]{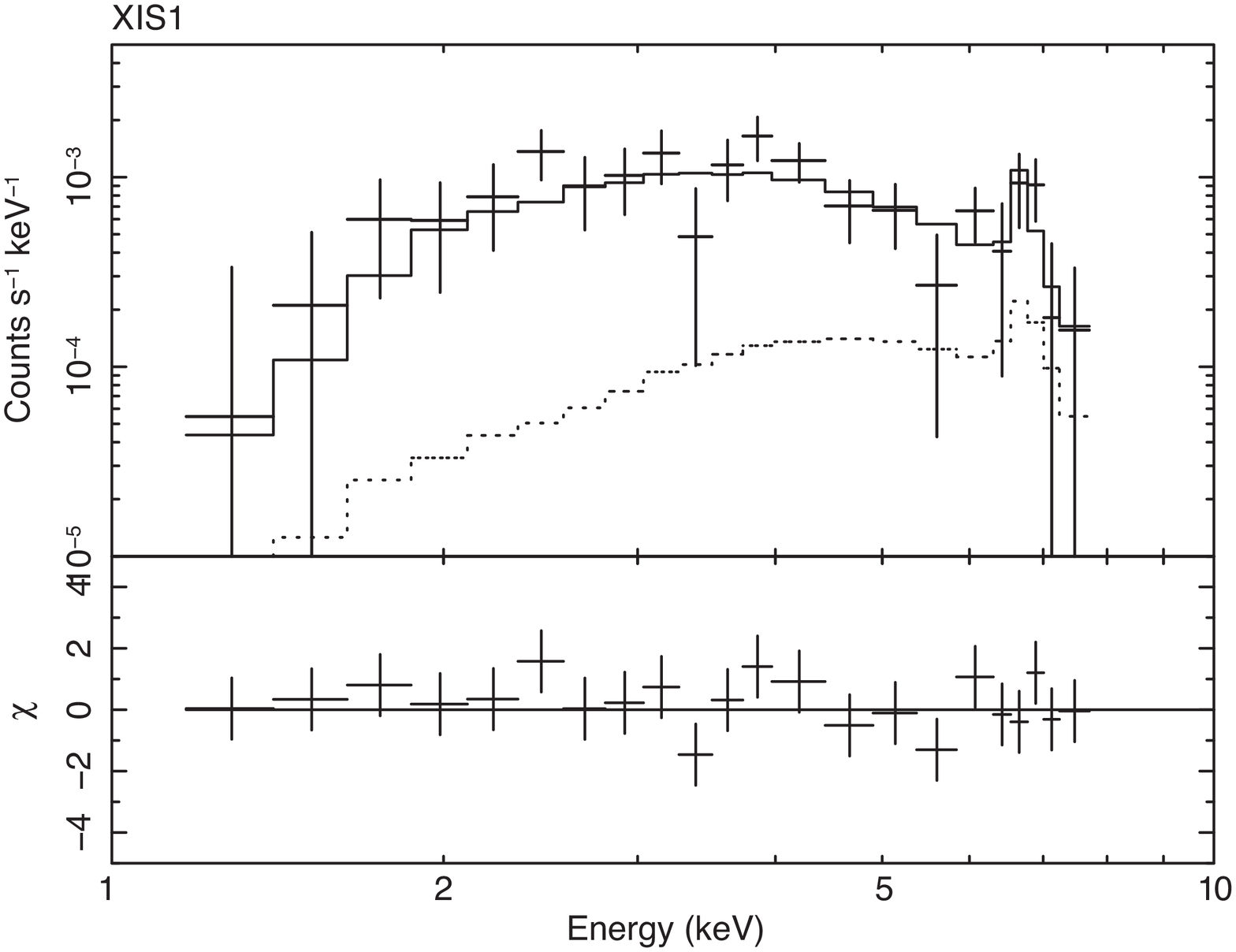}
\end{center}
\caption{X-ray spectra of Suzaku J174400$-$2913 (upper panel) and the residuals from the best-fit model (lower panel), 
left: XIS 0$+$3 and right: XIS 1.
The solid lines show the best-fit model, while the dotted lines show the contribution of the residual point source model.
}\label{fig:sample}
\end{figure*}


Since the photon numbers of the Chandra sources in the source region are limited to make a model spectra, less than 80 counts each, 
we made a model spectrum using  the integrated data of all the Chandra point sources with smaller than 80 net counts (Table 4 in \cite{Muno2004}). 
The normalization was adjusted to become the photon flux to be 7.0$\times$10$^{-6}$ photons s$^{-1}$ cm$^{-2}$ (2--8 keV).
The contribution of the remaining point sources to the source count rate in the 2--8 keV band was estimated to be 14--19 \%. 
The estimated spectrum of the remaining point sources is plotted in figure 2. 
Still we clearly see excess emission above the point source model.

We used the response files, Redistribution Matrix Files (RMFs) and Ancillary Response Files (ARFs) 
taken from {\tt xisrmfgen} and {\tt xissimarfgen}, respectively.
For the excess emission, we simultaneously fitted the XIS 0$+$3 and the XIS 1 spectra
with an absorbed power-law function. 
The cross sections of photoelectric absorption were taken from Morrison and McCammon (1983). 
This model was rejected with a 90\% confidence level (a $\chi^2$ value of 72.4 for d.o.f.=52), remaining large excess at $\sim 6.7$ keV.
Thus we added a narrow Gaussian line at 6.71($\pm$0.05) keV, 
then the fit was significantly improved with the $\Delta \chi^2$ value of 13.4, 
which is statistically highly significant of more than 99 \%  confidence level. 
The equivalent width was 850$^{+400}_{-390}$ eV.
Thus, we conclude that Suzaku J174400$-$2913 has a prominent 6.7 keV emission line, indicating a thin thermal origin.

Then we applied a thin thermal plasma model in a collisional ionization 
equilibrium state ({\tt apec} model in XSPEC) modified by low energy absorption.
The abundances  were  taken from Anders and Grevesse (1989).  
This model also gave an acceptable fit with $\chi^2$/ d.o.f.=59.2/51=1.16.
The best-fit model is plotted in figure 3.

Since the remaining point source photon flux has an error of 31\% (90\% confidence level),
we examined the best-fit parameters and their errors by changing the intensity (normalization) of 
the remaining point source spectrum  by $\pm$31\%.  
The results are listed in table 1.
The $N_{\rm H}$ value and the flux of Suzaku J174400$-$2913 in the 2--8 keV band are 
6.1$^{+2.5}_{-1.3}$$\times$10$^{22}$ cm$^{-2}$ and 2.5$\times10^{-13}$ erg s$^{-1}$ cm$^{-2}$, respectively.

\begin{table}
  \caption{Spectral parameters of Suzaku J174400$-$2913.}\label{tab:first}
  \begin{center}
    \begin{tabular}{lc}
      \hline
      Parameter 	& Value  \\
      \hline
      \multicolumn{2}{l}{Diffuse: {\tt wabs$\times$apec} } \\
      \hline 
        $N_{\rm H}$ ($\times$10$^{22}$ cm$^{-2}$) 	& 6.1$^{+2.5}_{-1.3}$  \\
        $kT$(keV)								&4.1$^{+2.7}_{-1.8}$  \\
        Abundance$^{\ast}$ (Solar)				& 0.58$^{+0.41}_{-0.32}$  \\
        Normalization$^{\dag}$ 					& 5.4$^{+5.1}_{-1.9}\times10^{-4}$\\
   \hline
$\chi^2$/d.o.f.  		& 59.2/51 \\
   \hline
    \end{tabular}
  \end{center}
$^{\ast}$ Relative to the solar value \citep{Anders1989}.\\
$^{\dag}$ The unit is 10$^{-14}$$\times VEM$/(4$\pi d^2$), 
where $VEM$ is the volume emission measure [cm$^{-3}$] and 
$d$ is the distance [cm].\\
\end{table}

\subsection{Time Variability}

In order to examine time variability, we made light curves in the 2--8 keV band from the source region for each detector.
After the background subtraction, the data of XIS 0, 1, and 3 were merged.
After the timing data were grouped for a 512 s bin, the light curve was fitted with a constant flux model. 
The $\chi^2$/d.o.f. value was 276/321=0.86, and hence no significant intensity variation was found.

\section{Discussion}

The best-fit  $N_{\rm H}$ value of 6.1$\times$10$^{22}$ cm$^{-2}$ is consistent with that of the GC
(e.g., \cite{Sakano2002}), and  hence we assume that Suzaku J174400$-$2913 is located near the GC, 
a 8 kpc distance from the Sun in the following discussion.

Suzaku J174400$-$2913 was stable in the short time scale during the Suzaku observation. 
The observed energy flux of Suzaku J174400$-$2913 is roughly consistent with that of the X-ray thread G359.55$+$0.16 \citep{Johnson2009}.  
Thus we see no noticeable time variability in the short and long time scale.
Furthermore, a power-law model fit for the XIS spectra gave a photon index of 2.3$^{+1.0}_{-0.8}$ and an $N_{\rm H}$ value of
6.5$^{+3.3}_{-2.1}\times10^{22}$ cm$^{-2}$, which are consistent with those from Chandra within the errors \citep{Johnson2009}.
These facts support that Suzaku J174400$-$2913 is a diffuse source and is associated with G359.55$+$0.16.
Contrary to \citet{Lu2003} and \citet{Johnson2009}, we concluded that Suzaku J174400$-$2913= G359.55$+$0.16 
is  thermal origin.
The Chandra spectrum had a lower signal-to-noise ratio and the data below 6.7 keV were used for the spectral analysis
\citep{Lu2003,Johnson2009}, which would be the reason that Chandra could not detect the 6.7 keV line.

Since the Suzaku point spread function is not good enough to determine the spatial structure, 
we assume the geometry of Suzaku J174400$-$2913 to be a cylinder 
with a diameter of 10$''$ and a length of 2$'$  \citep{Lu2003}.
Then the volume, $V$, was estimated to be 1.6$\times$10$^{55}$ cm$^3$.
Using the best-fit volume emission measure of  Suzaku J174400$-$2913, 
4.1$\times$10$^{56}$ cm$^{-3}$,
and $n_{\rm e}$=1.2$n_{\rm H}$, where $n_{\rm e}$ and $n_{\rm H}$ are the electron and hydrogen densities, respectively, 
we calculated the mean hydrogen and electron densities to be $n_{\rm H}$=4.6 cm$^{-3}$ 
and $n_{\rm e}$=5.5 cm$^{-3}$, respectively. 
The total thermal energy, $E_{\rm th}$, and the gas mass, $M$, were 
$E_{\rm th}$=(3/2)($n_{\rm H}+n_{\rm e}$)$kTV$=1.6$\times$10$^{48}$ erg and 
$M$=1.4$n_{\rm H} m_{\rm H}V$= 0.09M$_{\odot}$, respectively, 
where $k$, $T$ and $m_{\rm H}$ are the Boltzmann constant, the plasma temperature, and the hydrogen mass, respectively.

Chandra and XMM-Newton discovered many X-ray filamentary structures in the GC 
(e.g., Wang et al. 2002; Sakano et al. 2003; Lu et al. 2003, 2008; Johnson et al. 2009).
Among them, only G359.942$-$0.03 exhibited an emission line at $\sim6.7$ keV \citep{Johnson2009}.  
It has a point-like near-infrared source found in the Two-Micron All-Sky Survey (2MASS) \citep{Skrutskie2006} and 
a comet-like tail.  
\citet{Johnson2009} proposed that G359.942$-$0.03 is a ram pressure confined stellar wind bubble generated by a massive star.

Suzaku J174400$-$2913 is also filamentary but with no core. 
Thus it may not be stellar wind bubbles like G359.942$-$0.03.  
Since Suzaku J174400$-$2913 is aligned with the NTF G359.54$+$0.18 \citep{Yusef1997,Lu2003}, 
it may be a part of the GCXE confined by the magnetic field.
In fact, recent near infrared observations suggest that the GCXE could be confined by the magnetic field in the GC region \citep{Nishiyama2013}.
The thermal pressure,  $p_{\rm th}$, is calculated to be 
$p_{\rm th}$=($n_{\rm H}+n_{\rm e}$)$kT$$\sim$6.6$\times$10$^{-8}$ dyn cm$^{-2}$.
Assuming that the thermal pressure is equal to the  magnetic pressure, we can estimate the magnetic field strength to be 1.3 mGauss.

Related to the origin of the GCXE, we note here the other diffuse thermal sources with a $\sim$6.7 keV line in the GC region.  
Suzaku found an elongated source along the Galactic plane in the Sgr B2 region named G0.61$+$0.01 \citep{Koyama2007b}. 
They predicted that G0.61$+$0.01 is a part of a young SNR. 
From the filament-like morphology, an alternative scenario would be possible: 
G0.61$+$0.01 may also be a confined filament by the magnetic field, 
although the NTF has not been found near the source.  
The magnetic field is estimated to be $\sim$0.4--0.5 mGauss. 
These magnetic field strengths are comparable to those observed in the GC region, 0.1--1 mGauss (e.g., \cite{GCmagnetic1991,Morris1996}).
\citet{Senda2002} discovered the other thermal source with a strong Fe K-shell line, G0.570$-$0.018 using Chandra.
G0.570$-$0.018 has a ring-like structure with a radius of 10$''$ and a tail-like structure.  
They predicted to be a very young supernova remnant possibly in a free expansion phase.  
It was also detected with XMM-Newton and Suzaku \citep{Inui2009}, suggesting that the 6.4 keV line was variable, 
while the 6.7 keV line was stable.
The origin of the variable 6.4 keV line would be due to X-ray reflection nebula \citep{Koyama1996}, while
the stable 6.7 keV line and the plasma temperature of 6.1 keV are very similar to those of the GCXE.
Thus these thermal sources including Suzaku J174400$-$2913 would contribute to the GCXE.  
We suggest further deep survey observations may reveal many  thermal sources in the GC region. 

\section*{Acknowledgement}

The authors are grateful to all members of the Suzaku team. 
This work was supported by the Japan Society for the Promotion of Science (JSPS) 
KAKENHI Grant Numbers, 24540232 (S. Y.), 24740123 (M. N.), and 24540229 (K. K.).



\begin{thebibliography}{}
\bibitem[Anantharamaiah et al.(1991)]{GCmagnetic1991}
  Anantharamaiah, K. R., Pedlar, A., Ekers, R. D., \& Goss, W. M. 1991, \mnras, 249, 262
\bibitem[Anders \& Grevesse(1989)]{Anders1989}
   Anders, E., \& Grevesse, N. 1989, Geochim. Cosmochim. Acta, 53, 197
\bibitem[Baganoff et al.(2001)]{Baganoff2001}
   Baganoff, F. K., et al. 2001, \nat, 413, 45
\bibitem[Baganoff et al.(2003)]{Baganoff2003}
   Baganoff, F. K., et al. 2003, \apj, 591, 891
\bibitem[Eckart et al.(2002)]{Eckart2002}
   Eckart, A., Genzel, R., Ott, T., \& Sch\"{o}del, R. 2002, \mnras, 331, 917
\bibitem[Genzel \& Townes(1987)]{Genzel1987}
   Genzel, R., \& Townes, C. H. 1987, \araa, 25, 377
\bibitem[Genzel et al.(1994)]{Genzel1994}
   Genzel, R., Hollenbach, D., \& Townes, C. H. 1994, Rep. Prog. Phys., 57, 417
\bibitem[Genzel et al.(2000)]{Genzel2000}
   Genzel, R., Pichon, C., Eckart, A., Gerhard, O. E., \& Ott, T. 2000, \mnras, 317, 348
\bibitem[Ghez et al.(2000)]{Ghez2000}
   Ghez, A. M., Morris, M., Becklin, E. E., Tanner, A., Kremenek, T. 2000, \nat, 407, 349
\bibitem[Hyodo et al.(2008)]{Hyodo2008}
  Hyodo, Y., Tsujimoto, M., Hamaguchi, K., Koyama, K., Kitamoto, S., Maeda, Y., Tsuboi, Y., 
  \& Ezoe, Y. 2008, \pasj, 60, S85
\bibitem[Inui et al.(2009)]{Inui2009}
  Inui, T., Koyama, K., Matsumoto, H., \& Tsuru, T. G. 2009, \pasj, 61, S241
\bibitem[Johnson et al.(2009)]{Johnson2009}
   Johnson, S. P., Dong, H., \& Wang, Q. D. 2009, \mnras, 399, 1429
\bibitem[Koyama et al.(1989)]{Koyama1989}
   Koyama, K., Awaki, H., Kunieda, H., Takano, S., Tawara, Y., Yamauchi, S., 
   Hatsukade, I., \& Nagase, F. \nat, 1989, 339, 603
\bibitem[Koyama et al.(1996)]{Koyama1996}
   Koyama, K., Maeda, Y., Sonobe, T., Takeshima, T., Tanaka, Y., \& Yamauchi, S.
   \pasj, 1996, 48, 249
\bibitem[Koyama et al.(2007a)]{Koyama2007a}
   Koyama, K., et al.\ 2007a, \pasj, 59, S23
\bibitem[Koyama et al.(2007b)]{Koyama2007b}
   Koyama, K., et al. 2007b, \pasj, 59, S221
\bibitem[Koyama et al.(2007c)]{Koyama2007c}
   Koyama, K., et al. 2007c, \pasj, 59, S245
\bibitem[Lu et al.(2003)]{Lu2003}
   Lu, F. J., Wang, Q. D., \& Lang, C. C. 2003, \aj, 126, 319
\bibitem[Lu et al.(2008)]{Lu2008}
   Lu, F. J., Yuan, T. T., \& Lou, Y. -Q. 2008, \apj, 673, 915
\bibitem[Mitsuda et al.(2007)]{Mitsuda2007}
   Mitsuda, K., et al.\ 2007, \pasj, 59, S1
\bibitem[Morris \& Serabyn(1996)]{Morris1996}
   Morris, M., \& Serabyn, E. 1996, \araa, 34, 645
\bibitem[Morrison \& McCammon(1983)]{Morrison1983}
   Morrison, R., \& McCammon, D. 1983, ApJ, 270, 119
\bibitem[Muno et al.(2004)]{Muno2004}
   Muno, M. P., et al. 2004, \apj, 613, 1179
\bibitem[Muno et al.(2009)]{Muno2009}
   Muno, M. P., et al. 2009, \apjs, 181, 110
\bibitem[Nakajima et al.(2008)]{Nakajima2008}
   Nakajima, H., et al. 2008, \pasj, 60, S1
\bibitem[Nishiyama et al.(2013)]{Nishiyama2013}
   Nishiyama, S., et al. 2013, \apj, 769, L28
\bibitem[Ponti et al.(2010)]{Ponti2010}
   Ponti, G., Terrier, R., Goldwurm, A., Belanger, G., \& Trap, G. 2010, \apj, 714, 732
\bibitem[Ryu et al.(2013)]{Ryu2013}
   Ryu, S. G., Nobukawa, M., Nakashima, S., Tsuru, T. G., Koyama, K., \& Uchiyama, H. 2013, \pasj, 65, 33
\bibitem[Sakano et al.(2002)]{Sakano2002}
   Sakano, M., Koyama, K., Murakami, H., Maeda, Y., \& Yamauchi, S. 2002, \apjs, 138, 19
\bibitem[Sakano et al.(2003)]{Sakano2003}
   Sakano, M., Warwick, R. S., Decourchelle, A., \& Predehl, P. 2003, \mnras, 340, 747
\bibitem[Sch\"{o}del et al.(2002)]{Schoudel2002}
   Sch\"{o}del, R., et al. 2002, \nat, 419, 694
\bibitem[Senda et al.(2002)]{Senda2002}
   Senda, A., Murakami, H., \& Koyama, K. 2002, \apj, 565, 1017
\bibitem[Serlemitsos et al.(2007)]{Serlemitsos2007}
   Serlemitsos, P., et al.\ 2007, \pasj, 59, S9
\bibitem[Skrutskie et al.(2006)]{Skrutskie2006}
   Skrutskie, M. F., et al. 2006, \aj, 131, 1163
\bibitem[Tawa et al.(2008)]{Tawa2008}
   Tawa, N., et al. 2008, \pasj, 60, S11
\bibitem[Uchiyama et al.(2009)]{Uchiyama2009}
   Uchiyama, H., et al. 2009, \pasj, 61, S9
\bibitem[Uchiyama et al.(2013)]{Uchiyama2013}
   Uchiyama, H., Nobukawa, M., Tsuru, T. G., \& Koyama, K.
   2013, \pasj, 65, 19
\bibitem[Wang et al.(2002)]{Wang2002}
   Wang, Q. D., Gotthelf, E. V., \& Lang, C. C. 2002, \nat, 415, 148
\bibitem[Yamauchi et al.(1990)]{Yamauchi1990}
   Yamauchi, S. Kawada, M., Koyama, K., Kunieda, H., Tawara, Y., \& Hatsukade, I. 
   1990, \apj, 365, 532
\bibitem[Yusef-Zadeh et al.(1984)]{Yusef1984}
  Yusef-Zadeh, F., Morris, M., \& Chance, D., 1984, \nat, 310, 557
\bibitem[Yusef-Zadeh et al.(1997)]{Yusef1997}
  Yusef-Zadeh, F., Wardle, M., \& Parastaran, P., 1997, \apj, 475, L119
\end{thebibliography}
\end{document}